\documentclass[showpacs,preprintnumbers,eqsecnum]{revtex4}

\usepackage{amsmath}
\usepackage{makeidx}
\usepackage{graphicx}
\usepackage{bm}
\usepackage{hyperref}

\begin{document}

\preprint{}
\title{Analysis of the radiative thermal transfer in planar multi-layer systems with various emissivity and transmissivity properties}
\author{S. Spanulescu$^{1}$}
\affiliation{$^{1}$Department of Physics, Hyperion University of
Bucharest, 030615, Romania}

\begin{abstract}
\textbf{Abstract} The paper analyzes the radiative thermal transfer
in a liquid helium cryostat with liquid nitrogen shielding. A
infinite plane walls model is used for demonstrating a method for
lowering the radiative heat transfer and the numerical results for
two such systems are presented. Some advantages concerning the
opportunity of using semi-transparent walls are analytically and
numerically demonstrated.
\end{abstract}


\maketitle Keywords: Radiative heat transfer, emissivity, thermal
shield, cryostat
\section{Introduction}

\begin{figure*}[h]
\centering \centering
\includegraphics[width=4in]{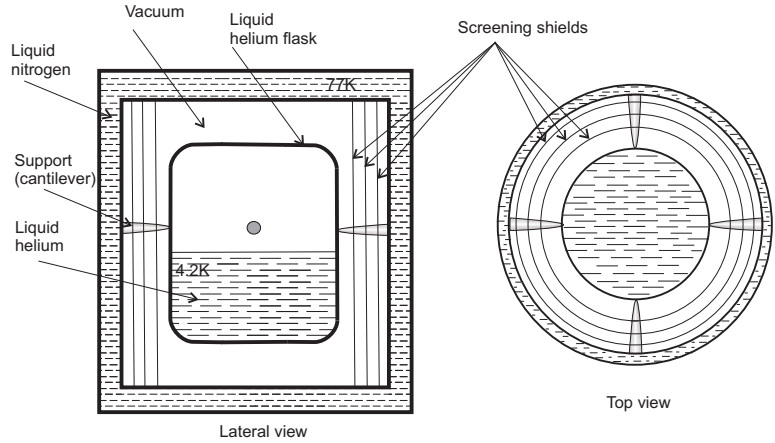}
\caption{The system of two infinite planes in vacuum}\label{fig0}
\end{figure*}

In a previous paper \cite{ssarxiv4} the thermal conductive flux in a
cryostat was analyzed, and some methods for its reduction were
suggested. In this paper we analyze the other major process of heat
transfer which occurs in the same system, the radiative heat
transfer. The classical method for lowering the values of this flux
involves a number of reflective screens interposed in the vacuum
chamber between the hot and the cold wall \cite{sieg}, as in figure
\ref{fig0}.

The net radiative thermal flux from the warm wall to the cold one is
strongly dependent on the properties of the walls and shields,
especially their emissivity \cite{hprs}, \cite{emis}. Although these
properties are not entirely known for cryogenic temperatures and
have to be experimentally determined, some theoretical consideration
concerning the general relationships describing the radiative
thermal transfer may be drawn and they allow the proper design for
improved cryostats.

We analyze in the following the basic relationships for calculating
this process, and a method for significantly reducing it is
presented.

\section{Radiative heat transfer between reflective infinite planes in vacuum}

\begin{figure*}[h]
\centering \centering
\includegraphics[width=1.5in]{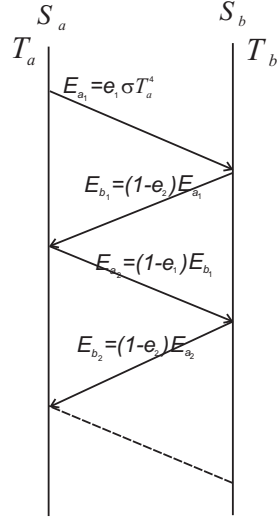}
\caption{The system of two infinite planes in vacuum}\label{fig1}
\end{figure*}

Let us consider the system composed by two plane surfaces ${S_a}$
and ${S_b}$ placed in vacuum as in figure \ref{fig1} in steady state
conditions at temperatures ${T_a}$ and ${T_b}$.

We analyze the radiative thermal flux between the two surfaces, from
${S_a}$ to ${S_b}$  if the first one is in contact with liquid
nitrogen at the temperature ${T_a} = 77.36\rm K$ and the second one
is in contact with liquid helium at the temperature ${T_b} = 4.22\rm
K$.

The thermal flux emitted by the surface unit by ${S_a}$, if it
presents the emissivity ${e_1}$ considered constant with the
temperature is (figure \ref{fig1}):
    \begin{equation}
    {E_{{a_1}}}= {e_1}\sigma T_a^4  \end{equation}
conforming to the Stephan-Boltzmann law, with the constant
\cite{landau}  $\sigma = \frac{{2{\pi ^2}k_B^4}}{{15{h^3}{c^2}}} =
5,6692 \cdot {10^{ - 8}}\rm W/({m^2}{K^4})$

This thermal flux is partially reflected by the surface ${S_b}$, so
that a fraction of it is returned to the surface ${S_a}$. According
to Kirchhoff  law, the reflexivity $r$ of the surface is $r=1-e$, so
that this fraction is \cite{chis}
      \begin{equation}{E_{b1}} = (1 - {e_2}){E_{a1}} = {e_1}(1 - {e_2})\sigma T_a^4\end{equation}
This flux is again reflected to ${S_a}$,as a secondary flux
${E_{{a_2}}}$.
    \begin{equation}{E_{{a_2}}} = (1 - {e_1}){E_{{b_1}}} = {e_1}(1 - {e_1})(1 - {e_2})\sigma T_a^4\end{equation}

    After other two reflections the surface ${S_a}$ transmits also the flux:

    \begin{equation}{E_{{a_3}}} = {e_1}{(1 - {e_1})^2}{(1 - {e_2})^2}\sigma T_a^4\end{equation}
and after  $k$ double reflections the flux is:

    \begin{equation}{E_{{a_k}}} = {e_1}{(1 - {e_1})^{(k - 1)}}{(1 - {e_2})^{(k - 1)}}T_a^4\end{equation}

It follows that the total flux ${E_0}$ emitted by ${S_a}$ and
incident on ${S_b}$ is:

    \begin{equation}\begin{gathered}
  {E_a} = {E_{a1}} + {E_{a2}} + ... + {E_{ak}}+... = {e_1}\sigma T_a^4 + {e_1}(1 - {e_1})(1 - {e_2})\sigma T_a^4 \hfill \\
   + {e_1}{(1 - {e_1})^2}{(1 - {e_2})^2}\sigma T_a^4 + ... + {e_1}{(1 - {e_1})^{k-1}}{(1 - {e_2})^{k-1}}\sigma T_a^4+... \hfill \\
\end{gathered} \end{equation}
    \begin{equation}{E_a} = {e_1}\sigma T_a^4\frac{1}{{1 - (1 - {e_1})(1 - {e_2})}}\end{equation}

    The total flux ${E_{ab}}$ emitted by ${S_a}$ and absorbed by  ${S_b}$ is:
    \begin{equation}{E_{ab}} = {e_2}{E_a} = {e_2}{e_1}\sigma T_a^4\frac{1}{{1 - (1 - {e_1})(1 - {e_2})}}\label{1.8}\end{equation}

\begin{figure*}[h]
\centering \centering
\includegraphics[width=2in]{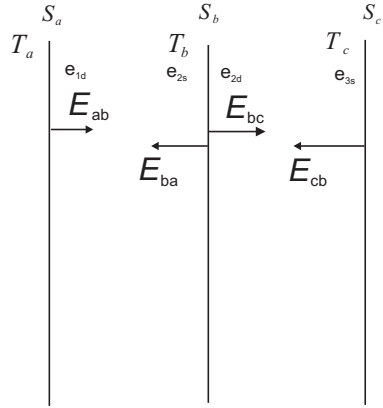}
\caption{The system of two infinite planes in vacuum}\label{fig2}
\end{figure*}

For calculating the net flux flowing from the wall ${S_a}$ to the
wall ${S_b}$ we must subtract the flux emitted by ${S_b}$ and
absorbed by ${S_a}$ (as in  figure \ref{fig2}). At the temperature
${T_b}$ this flux has an expression similar with (\ref{1.8}):
    \begin{equation}{E_{ba}} = \frac{{{e_1}{e_2}\sigma T_b^4}}{{1 - (1 - {e_1})(1 - {e_2})}}\label{1.8a}\end{equation}

    If ${T_a} > {T_b}$, the thermal flux emitted by ${S_a}$ is greater than that emitted by ${S_b}$, and the total net thermal flux from ${S_a}$ to ${S_b}$ is:
    \begin{equation}E = {E_{ab}} - {E_{ba}}\end{equation}
    \begin{equation}E = \frac{1}{{\frac{1}{{{e_1}}} + \frac{1}{{{e_2}}} - 1}}\sigma (T_a^4 - T_b^4)\label{1.8b}\end{equation}

    In the case that the walls have the same emissivity ${e_1} = {e_2} = e$:
    \begin{equation}E = \frac{e}{{2 - e}}\sigma (T_a^4 - T_b^4)\end{equation}

For calculating the net thermal flux the emissivity coefficients
${e_1}$ and ${e_2}$ and the temperatures are necessary. If the
temperature ${T_b}$ is not given, it may be calculated taking into
consideration that if ${S_b}$ is placed in vacuum in steady state
conditions the thermal flux absorbed is equal with the emitted one.
So, imposing a value for the total net flux through the surface
${S_b}$,its temperature is:
    \begin{equation}{T_b} = {\left[ {T_a^4 - \frac{E}{\sigma }\left( {\frac{1}{{{e_1}}} + \frac{1}{{{e_2}}} - 1} \right)} \right]^{\frac{1}{4}}}\end{equation}

In the case that between the two walls there is a third wall with
the left side emissivity $e_{2s}$ and the right side emissivity
$e_{2d}$ (as in figure \ref{fig2}), for calculating the thermal flux
we have to consider the condition that in steady state regime the
thermal flux emitted by  ${S_a}$ and absorbed by ${S_b}$ is equal to
the flux emitted by ${S_b}$ and absorbed by ${S_c}$. We obtain the
equations
    \begin{equation}\left\{ \begin{gathered}
  E = \frac{1}{{\frac{1}{{{e_{1d}}}} + \frac{1}{{{e_{2s}}}} - 1}}\sigma (T_a^4 - T_b^4) \hfill \\
  E = \frac{1}{{\frac{1}{{{e_{2d}}}} + \frac{1}{{{e_{3s}}}} - 1}}\sigma (T_b^4 - T_c^4) \hfill \\
\end{gathered}  \right.\end{equation}

If the exterior walls temperatures ${T_a}$ and ${T_c}$ are known, by
solving this system we obtain the unknowns $E$ and ${T_b}$
    \begin{equation}E = \frac{1}{{\frac{1}{{{e_{1d}}}} + \frac{1}{{{e_{2s}}}} + \frac{1}{{{e_{2d}}}} + \frac{1}{{{e_{3s}}}} - 2}}\sigma (T_a^4 - T_c^4)\end{equation}
    \begin{equation}{T_b} = {\left[ {\frac{{\left( {\frac{1}{{{e_{3s}}}} + \frac{1}{{{e_{2d}}}} - 1} \right)T_a^4 + \left( {\frac{1}{{e2}} + \frac{1}{{{e_{1d}}}} - 1} \right)T_c^4}}{{\frac{1}{{{e_{3s}}}} + \frac{1}{{{e_{2s}}}} + \frac{1}{{{e_{1d}}}} + \frac{1}{{{e_{2d}}}} - 2}}} \right]^{\frac{1}{4}}}\end{equation}

In the case that all the surfaces have the same emissivity ${e_{1d}}
= {e_{2s}} = {e_{2d}} = {e_{3s}} = e$, the expressions simplify as

    \begin{equation}E = \frac{e}{{2(2 - e)}}\sigma (T_a^4 - T_c^4)\end{equation}
    \begin{equation}{T_b} = {\left( {\frac{{T_a^4 + T_c^4}}{2}} \right)^{\frac{1}{4}}}\end{equation}

One may notice that the thermal flux is twice smaller that in the
case of two walls, and the fourth power of middle wall temperature
is the average value of the exterior temperature at the fourth
power.

Generalizing, for the case of $n$ walls with the same emissivity
$e$, and with the extreme temperatures $T_1$ and $T_2$, the total
net flux transmitted from the warmest wall and absorbed by the
coldest one is

    \begin{equation}E = \frac{e}{{(n - 1)(2 - e)}}\sigma (T_1^4 - T_n^4)\label{nwalls}\end{equation}
and the intermediate walls temperatures $T_k$ are
    \begin{equation}{T_k} = {\left\{ {\frac{1}{n-1}[(n - k)T_1^4 + (k - 1)T_n^4]} \right\}^{\frac{1}{4}}},\quad k=2,3,...,n-1\end{equation}

In the table \ref{tab1} we present numerical results for the
radiative heat transfer between two plane infinite walls with
different emissivity coefficients. One may notice that the minimum
radiative heat transfer is obtained for the lowest values of the
emissivity of both walls, as expected. These numerical results may
be used as a reference for more sophisticated methods for reducing
the thermal flux, as it will be presented further. For a system of
$n$ walls with the same emissivity, the corresponding value on the
diagonal should be divided by $n-1$.

\begin{table*}[hb]
\caption{\label{tab1}  The heat transfer power reported to the
surface between two plane walls, for various values of the
emissivity of the right side and the left side ($\rm W/m^2$).}
\footnotesize\rm
\begin{tabular}{l |l l l l l l l l l l l}
\hline
 $_{e_l} \backslash^ {e_r}$& 0.1 & 0.2 & 0.3 & 0.4 & 0.5 & 0.6 & 0.7 & 0.8 & 0.9 \\
 \hline

 0.1 & 0.105 & 0.142 & 0.162 & 0.173 & 0.181 & 0.187 & 0.191 & 0.194 & 0.197 \\
 0.2 & 0.142 & 0.221 & 0.272 & 0.307 & 0.332 & 0.352 & 0.367 & 0.38 & 0.39 \\
 0.3 & 0.162 & 0.272 & 0.352 & 0.412 & 0.46 & 0.498 & 0.53 & 0.556 & 0.579 \\
 0.4 & 0.173 & 0.307 & 0.412 & 0.498 & 0.569 & 0.629 & 0.68 & 0.725 & 0.763 \\
 0.5 & 0.181 & 0.332 & 0.46 & 0.569 & 0.664 & 0.747 & 0.821 & 0.886 & 0.944 \\
 0.6 & 0.187 & 0.352 & 0.498 & 0.629 & 0.747 & 0.854 & 0.951 & 1.04 & 1.121 \\
 0.7 & 0.191 & 0.367 & 0.53 & 0.68 & 0.821 & 0.951 & 1.073 & 1.187 & 1.294 \\
 0.8 & 0.194 & 0.38 & 0.556 & 0.725 & 0.886 & 1.04 & 1.187 & 1.329 & 1.464 \\
 0.9 & 0.197 & 0.39 & 0.579 & 0.763 & 0.944 & 1.121 & 1.294 & 1.464 &
 1.631\\

\hline
\end{tabular}
\end{table*}

\begin{figure*}[h]
\centering \centering
\includegraphics[width=5in]{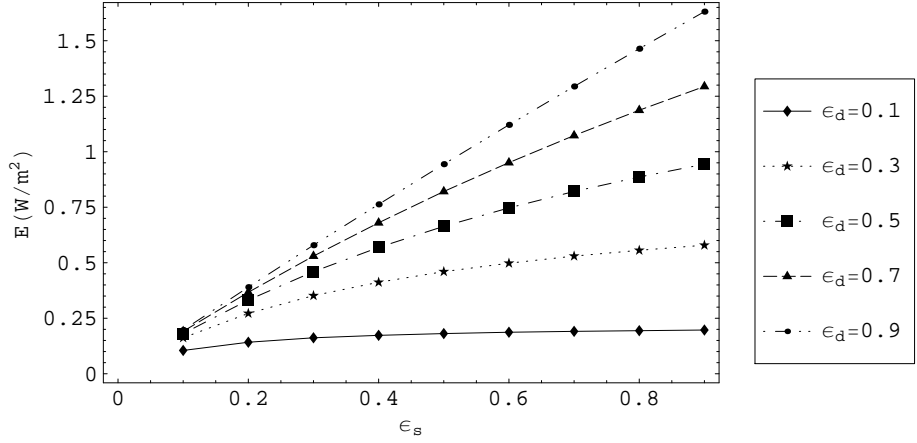}
\caption{The amount of heat transfer by radiation in the system of
two infinite planes in vacuum}\label{fig3}
\end{figure*}

\section{Radiative heat transfer in the case of semi-transmissive
walls}

If a part $T$ of the incident thermal radiation $I$ passes through
the wall (supposed thermic "thin") and another part $R$ is returned
to the region where it came, it may be possible that, in certain
condition, the total net heat flux to be diminished.

In steady state conditions we may write
\begin{equation}
I=T+R+A
\end{equation}
where $A$ is the radiation amount that is absorbed by the wall.

If we denote by
\begin{equation}
t=\frac{T}{I},\;r=\frac{r}{I},\;e=\frac{A}{I}
\end{equation}
the transmissivity, reflexivity and emissivity of the wall, taking
into account the Kirchhoff low of radiation which states that the
emitted radiation is equal with the absorbed one, we obtain
\begin{equation}
t=1-e-t
\end{equation}

Therefore the equations (\ref{1.8}) and (\ref{1.8a}) have to be
written with this reflection coefficient instead $1-e$, so that the
net flux emitted by the semi-transparent wall $S_a$ and absorbed by
$S_b$ is
  \begin{equation}{E_{st}}= {e_2}{e_1}\sigma \frac{T_a^4-T_b^4}{{1 - (1 - {e_1}-t_1)(1 - {e_2}-t_2)}}\label{2.1}\end{equation}
If we use a wall with a transparency close to
\begin{equation}
t=1-e \label{2.2}
\end{equation}
(an ideal case of semi-transparency) then the denominator in the
relationship (\ref{1.8b}) becomes unity, and the thermal flux
emitted by this wall and absorbed by the other becomes
   \begin{equation}{E_{ab}} = {e_2}{e_1}\sigma\left(T_a^4-T_a^4\right)\label{2.0}\end{equation}
which reduces the heat transfer by the ratio
\begin{equation}
\frac{E}{E_{st}}=\frac{1}{{1 - (1 - {e_1})(1 - {e_2})}}
\end{equation}

For example, with a typical high reflective surface with $e=0.1$, if
the transmissivity is $t=0.9$, the heat transfer reduction ratio is
almost five.

However, for heat isolation purposes one may take into account the
necessity that the transmitted part of the flux may still be
absorbed after the final, coldest wall. That is why the walls must
have an unidirectional transparency, from the colder wall to the
warmer one. This way, the thermal radiation is reflected to the
exterior of the system and dissipated there.

Hence, the right formula for calculating the heat transfer by
radiation is obtained considering $t_2=0$ in the expression
(\ref{2.1})

\begin{equation}{E_{st}}  = {e_2}{e_1}\sigma T_a^4\frac{1}{{1 - (1 - {e_1}-t_1)(1 - {e_2})}}\label{2.3}\end{equation}

In the ideal case given by the condition (\ref{2.2}), the net
thermal flux is given by equation (\ref{2.0}), being considerably
lower than in the opaque wall case.

If there are $n$ walls with the same emissivity $e$ and one side
transparency $t=1-e$, the net thermal flux is obtained with an
expression similar to (\ref{nwalls})
  \begin{equation}E_{st} = \frac{e}{n - 1}\sigma (T_1^4 - T_n^4)\label{nwalls}\end{equation}

\section{Numerical results and conclusions}

\begin{figure*}[h]
\centering \centering
\includegraphics[width=5in]{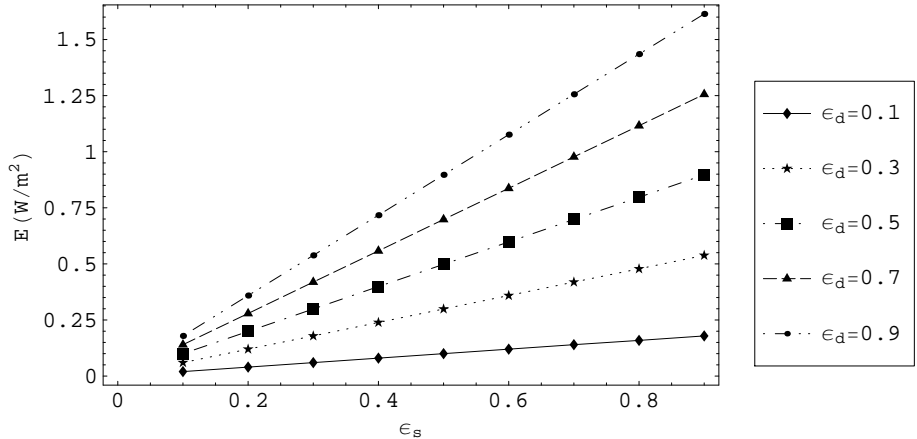}
\caption{The amount of heat transfer by radiation in the system of
two infinite planes in vacuum}\label{fig4}
\end{figure*}

The numerical results obtained for a system of two walls with ideal
transparency (the warmer one) are presented in table \ref{tab2} and
figure \ref{fig4}. One may notice an important reduction of the
radiation flux especially for the common case of low emissivity
values.

\begin{table*}[hb]
\caption{\label{tab2}  The heat transfer power reported to the
surface between two plane ideally semi-transparent walls, for
various values of the emissivity of the right side and the left side
($\rm W/m^2$).} \footnotesize\rm
\begin{tabular}{l |l l l l l l l l l l l}
\hline
 $_{e_l} \backslash^ {e_r}$& 0.1 & 0.2 & 0.3 & 0.4 & 0.5 & 0.6 & 0.7 & 0.8 & 0.9 \\
 \hline

 0.1 & 0.02 & 0.04 & 0.06 & 0.08 & 0.1 & 0.12 & 0.14 & 0.159 & 0.179 \\
 0.2 & 0.04 & 0.08 & 0.12 & 0.159 & 0.199 & 0.239 & 0.279 & 0.319 & 0.359 \\
 0.3 & 0.06 & 0.12 & 0.179 & 0.239 & 0.299 & 0.359 & 0.419 & 0.478 & 0.538 \\
 0.4 & 0.08 & 0.159 & 0.239 & 0.319 & 0.399 & 0.478 & 0.558 & 0.638 & 0.717 \\
 0.5 & 0.1 & 0.199 & 0.299 & 0.399 & 0.498 & 0.598 & 0.698 & 0.797 & 0.897 \\
 0.6 & 0.12 & 0.239 & 0.359 & 0.478 & 0.598 & 0.717 & 0.837 & 0.957 & 1.076 \\
 0.7 & 0.14 & 0.279 & 0.419 & 0.558 & 0.698 & 0.837 & 0.977 & 1.116 & 1.256 \\
 0.8 & 0.159 & 0.319 & 0.478 & 0.638 & 0.797 & 0.957 & 1.116 & 1.275 & 1.435 \\
 0.9 & 0.179 & 0.359 & 0.538 & 0.717 & 0.897 & 1.076 & 1.256 & 1.435 &
 1.614\\
\hline
\end{tabular}
\end{table*}

Of course, there are technological difficulties concerning the
obtaining of the right condition of transparency, but it may be
achieved using thin film deposition of materials with the proper
index of refraction. One may use the Fresnel formulae for the
reflective coefficient calculus, but it has to be considered the
temperature and especially the wavelength dependence of the index of
refraction of the involved materials, and some other special effects
mentioned in the literature \cite{Robi}.

The presented formulae may be used as a first step in evaluating the
performances of cryogenic shields, but surely some experimental
steps have to be included for a final cryostat design.

\section*{Acknowledgements} This work was supported by the Romanian National
Research Authority (ANCS) under Grant 22-139/2008.

\end{document}